\documentclass[12pt]{article}
\usepackage{graphicx}
\usepackage{url}

\def\Title#1{\begin{center} {\Large #1 } \end{center}}
\def\Author#1{\begin{center}{ \sc #1} \end{center}}
\def\Address#1{\begin{center}{ \it #1} \end{center}}

\newenvironment{Abstract}{\begin{quotation}  }{\end{quotation}}
\newenvironment{Presented}{\begin{quotation} \begin{center} 
             PRESENTED AT\end{center}\bigskip 
      \begin{center}\begin{large}}{\end{large}\end{center} \end{quotation}}

\def\beq{\begin{equation}}
\def\eeq#1{\label{#1}\end{equation}}
\def\eeqn{\end{equation}}

\def\beqa{\begin{eqnarray}}
\def\eeqa#1{\label{#1}\end{eqnarray}}
\def\eeqan{\end{eqnarray}}

\let\bar=\overbar

\def\Dslash{\not{\hbox{\kern-4pt $D$}}}
\def\dslash{\not{\hbox{\kern-2pt $\del$}}}

\def\msb{{\bar{\ssstyle M \kern -1pt S}}}

\begin{document}
\begin{titlepage}

\vfill

\Title {Summary of the recent PHYSTAT-$\nu$ Workshops}
\vfill
\Author{Louis Lyons}
\Address{Blackett Lab, Imperial College, London SW7 2BK, UK \\
and\\
Particle Physics, Oxford OX1 3RH, UK}
\vfill
\begin{Abstract}
This is a summary of the recent PHYSTAT-$\nu$ Workshops in Japan and at Fermilab, on `Statistical
Issues in Experimental Neutrino Physics'.
\end{Abstract}
\vfill
\begin{Presented}
NuPhys2016, Prospects in Neutrino Physics 

Barbican Centre, London UK, December 12--14, 2016
\end{Presented}
\vfill
\end{titlepage}
\def\thefootnote{\fnsymbol{footnote}}
\setcounter{footnote}{0}


\section{The PHYSTAT Workshop series}
The PHYSTAT series of Workshops deals with the {\bf statistica}l issues that arise in analyses in
High Energy Physics (and sometimes in Astroparticle Physics). The first two were devoted to the
topic of Upper Limits and took place at CERN and at Fermilab in 2000. Since then Worshops have 
been held at Durham (2002), SLAC (2003), Oxford (2005), and CERN again (2007 and 2011). Information
about these can be traced back via ref.  \cite{CERN2011}.

There had not been much participation by neutrino physicists, so it was decided that there should 
be meetings devoted specifically to the issues that arise in analysing the results of neutrino 
experiments, one at IPMU in Kashiwa, Japan, and the other at Fermilab;
these attracted about 90 and 130 participants respectively. More detailed information is available on their
web-sites\cite{nu_IPMU,nu_FNAL}.

\section{PHYSTAT-$\nu$ Programmes} 
The programmes of the two meetings were similar. They consisted of:
\begin{itemize}
\item{Introductory Statistical Material (see Section \ref{Introduction}).}  
\item{Summary of  Neutrino Physics: What to measure}
\item{Invited and Contributed talks, including talks by Statisticians}
\item{Panel discussion}
\item{Poster session}
\item{Summary talks by a Physicist and by a Statistician (Bob Cousins and David van Dyk at IPMU, Asher Kaboth 
and Richard Lockhart at Fermilab.)}
\end{itemize}

In addition, at Fermilab there was a talk on the statistical issues involved in the recent discovery of gravitational waves 
by the LIGO Collaboration.

\section{Introductory Topics}
\label{Introduction}
This was a brief summary providing information at a simple level on some of the topics that would be discussed
more deeply during the rest of the meeting. They included
\begin{itemize}
       \item{Combining results of different analyses: 
For correlated parameters, this can sensibly produce best values outside the range of the individual values, and greatly 
reduced uncertainties.}
       \item{Coverage:
This is a property of statistical procedures for Parameter Determination. For a repeated series of measurements, it is the fraction 
of parameter ranges that contain the true value.}
       \item{Blind Analysis:
Various methods are available to prevent the Physics result being known until the analysis is complete. This reduces the 
possibility of the Physicist subconsciously biassing the result in their preferred prejudice.}
       \item{$p$-value: 
This is the probability of obtaining a measurement at least as discrepant as ours, assuming some hypothesis. They are
widely misunderstood as the probability of the hypothesis betowardsg true, given the data.}
       \item{Significance:
$p$-values are commonly converted into significance (i.e number of $\sigma$s), assuming a single-sided Gaussian tail area.} 
       \item{Combining $p$-values:
Their is no unique way of doing this.}
      \item{ Upper Limits on cross-sections, etc: These can be useful in excluding models.}
      \item{ LEE = Look Elsewhere Effect: A peak in a spectrum can be due to exciting New Physics or to a boring fluctuation. 
The probabiity of this occuring anywhere in an analysis is larger than that for a fluctuation at the position seen in the data.}
       \item{Why $5\sigma$ for discovery?
See Section \ref{5sigma}.}
       \item{Comparing 2 hypotheses:
Examples include the Neutrino Mass Hierarchies; whether sterile neutrinos exist; etc.}
       \item{Wilks’ Theorem: 
See Section \ref{Wilks}.}
\end{itemize}

\section{Physics Topics}
Within the realm of neutrino physics, subjects for which statistical issues seem particularly relevant and which produced 
interesting discussions included:
\begin{itemize} 
\item{Fitting parameters for 3 neutrino oscillation situations}
\item{Searching for sterile neutrinos}
\item{Determining the neutrino mass hierarchy}
\item{Determining the CP phase}
\item{Searching for rare processes, e.g. ultra high energy cosmic neutrinos, neutrino-less double beta 
decay\footnote{Although this decay process has no neutrinos, they are involved virtually, and the decay rate could provide
information on  neutrino properties.}, supernovae neutrinos, etc.}
\item{Neutrino cross-sections}
\item{Reconstruction and classification issues, e.g. for rings in Cerenkov detectors} 
\end{itemize}

\section{Statistical issues}
\label{Statistical}
\subsection{Multi-variate techniques}
These are widely used in data selection e,g. for preferentially rejecting signal compared with background. Typical techniques are 
boosted decision trees, neural networks, etc. 
There is no need to regard neural networks as `black boxes', as it is easy to understand how a network with one hidden layer
operates. At the FNAL meeting, there was a talk on Deep Learning, which uses a neural network 
with many  layers of nodes, which are supposed to provide better discrimination.

With any multivariate method, it is important to assess its properties, including sensitivity to systematics; and to ensure that
the training events cover the region of phase space that the data occupy. 

\subsection{Treatment of systematics}
In almost all analyses, the expected distribution of data depends not only on the parameter of interest $\phi$ 
(e.g. the neutrino mass hierarchy), but also on so-called nuisance parameters $n$; they could be other interesting 
physics parameters (e.g. the CP phase) or various experimental systematics (e.g. the energy scale). To quote 
a result for $\phi$ requires some procedure for dealing with the nuisance parameter(s). 

One possibility is to quote a range for $\phi$ for each value of $n$.  This has the advantage that if subsequently the 
knowledge about $n$ is improved, we can easily incorporate this to obtain an improved range for $\phi$. An alternative 
is to eliminate $n$ by profiling or marginalistation. The former requires calculating the best value of $n$ ($n_{best}(\phi))$
for each value of $\phi$.
It is sensible to apply this to a likelihood function; then the profile likelihood is given by
\begin{equation}
{\mathcal L}_{prof}(\phi) = {\mathcal L}(\phi,n_{best}(\phi))
\end{equation}  
On the other hand, marginalisation involves integrating over $n$. It is used to convert the Bayesian 
posterior probability density  $p(\phi, n)$ into one just for $p(\phi)$:
\begin{equation}
p(\phi) = \int p(\phi,n)\  dn
\end{equation}
For situations where the likelihood function ${\mathcal L}(\phi,n)$ is a two (or more) dimensional Gaussian and the priors are uniform,
marginalisation and profiling lead to the same result.

\subsection{Non-asymptotic behaviour}
With enough data, asymptotic approximations can sometimes be used to simplify an analysis. For example the log-likelihood ratio for two hypotheses  $-2\ln({\mathcal L}_0/{\mathcal L}_1)$  involves summations over the observed events, and so by the Central Limit
Theorem (CLT) should become Gaussian distributed. 

Another example is the expected number of degrees of freedom when fitting the
2-neutrino flavour survival probability $P = 1- \sin^2 2\theta\ \sin^2 (\Delta m^2 L/E)$ to a lepton energy spectrum; usually 
there are two free parameters $\sin^2 2\theta$ and $\Delta m^2$, but for small  $\Delta m^2 L/E$, the data are sensitive only
to the combination   $\sin^2 2\theta (\Delta m^2)^2$, unless there is really a lot of data.  However, neutrino experiments
sometimes have limited statistics, and so the approximation may not be valid . It is then necessary to determine the expected distribution, usually by  Monte Carlo simulation.

\subsection{Unphysical parameter values} 
There are often fierce arguments in large collaborations as to whether quoted values
for physical parameters with well-defined ranges (e.g. $\sin^2 2\theta$, or the mass of a particle) should 
be confined to their physical ranges. The answer, of course, depends on how the result is to be used.

\subsection{Bayes or Frequentism?}
The `Bayes versus Frequentism' choice is often the cause for intense discussion. 
At the Kashiwa meeting Steve Biller gave a vigorous critique of frequentist approaches.

 In other fields, Baysianism tends to be far 
more used than in Particle Physics. At the LHC, using both Bayesian and Frequentist methods for measuring a given parameter 
is regarded as desirable. However, Bayesian methods are not recommended for hypothesis testing (i.e. comparing data with 
different hypotheses)  because of the stronger depedence on the choice of the Bayesian prior.

Even for parameter determination, the choice of prior can be non-trivial. For example, should
we express our ignorance about the CP phase angle by choosing a prior uniform in angle, or in its sine, or
some other functional form?

\subsection{Why $5\sigma$ for discovery?}
\label{5sigma}
For collider experiments, the standard criterion for claiming a discovery involves the $p$-value for the null 
hypothesis (i.e. no New Physics) $p_0$ being less than $3\ 10^{-7}$, equivalent to $5\sigma$. Reasons include past 
experience with incorrect discovery claims at lower levels; 
the look-elsewhere effect; underestimated systematics; and the fact that `extraordinary claims require 
extraordinary evidence'. Not all analyses are equally affected by the last three points so there is an 
argument in favour of having  a variable discovery criteron, but there clearly are problems in implementing that. 

\subsection{Why $CL_s$ for exclusion?}
Exclusion of $H1$, an alternative hypothesis involving new physics, usually requires its $p$ value ($p_1$) to 
be smaller than, say, 0.05. In collider experiments, however,  the variable often used is $CL_s = p_1/(1 - p_0)$.
This is to provide protection against the $5\%$  chance of excluding $H1$ even when an analysis has little or no
sensitivity to it. Thus $CL_s$ is a conservative modification of a frequentist procedure.

\subsection{Wilks Theorem} 
\label{Wilks}
If we compare our data with two hypotheses $H0$ and $H1$, we can use the difference
in the two $\chi^2$ (or almost equivalently -2 times the ln-likelihood ratio) to judge which hypothesis better 
explains the data. This could apply for:\\
(a) Using a straight line or a quadratic form to fit some data. \\
(b) A mass spectrum. $H0$ could be a background only distribution, while $H1$ could be background plus
signal, with the signal parametrised by its mass $M$ and production rate $\mu$. \\
(c) $H0$ and $H1$ could be the normal and inverted neutrino mass hierarchies. \\
For all these cases, if $\chi^2_1$ for $H1$ is much smaller than $\chi^2_0$ for $H0$, we would generally accept $H1$
in preference to $H0$. Wilks Theorem gives a way of judging whether or not the difference $\Delta\chi^2 = \chi^2_0
- \chi^2_1$ is small. It applies provided \\
(i) $H0$ is true \\
(ii) The hypotheses are nested i.e. $H1$ can be reduced to $H0$ by setting to special values (e.g. zero) any extra 
parameters in $H1$ but not in $H0$ \\
(iii) The extra parameter values to achieve this are all well-defined and not on the boundaries of their
allowed ranges. \\
(iv) There is enough data for assymptotic approximations to be valid. \\
 Thus the theorem applies to situation (a); then $\Delta \chi^2$ should be distributed as $\chi^2$ with the number of degrees of 
freedom equal to the number of extra parameters in $H1$. However, it does not apply to (b) (because when $\mu = 0$, 
$M$ is irrelevant) or to (c) ($H0$ and $H1$ are not nested). Even when the theorem does not apply, $\Delta\chi^2$ can still be a useful variable, but its expected distribution must then be determined, usually by simulation.

\subsection{Neutrino Mass Hierarchy}
For comparing `simple' hypotheses (ones with no free parameters), the Neyman-Pearson lemma\cite{NP} states that the likelihood ratio is the best  data statistic for separating the hypotheses. The situation does not quite apply here, because of experimental
nuisance parameters, and also because of uncertainties in the values of other relevant physics parameters (e.g. the CP 
phase). However $q = -2 \ln ({\mathcal L}_{NH}/{\mathcal L}_{IH})$ is still likely to be a useful variable, and is used as the data statistic.
As already mentioned, by the CLT the distributions of  $q$ under the two hypotheses are asymptotically Gaussian, but 
the Gaussians are often taken to be approximately symmetrically situated at $\pm T$ and with equal widths $2\sqrt T$ (see,
for example, references \cite{Q,B}). Whether this is so in particular circumstances needs to be checked by simulation. 
There are certainly similar Physics examples where it is not so.  
 
\subsection{Combining results (e.g. cross-sections) with unknown correlations}
When the correlations are unknown, it is impossible to combine different results optimally.
Assuming that there are no correlations is not always sensible or conservative.

\subsection{Unfolding}
In comparing experimental distributions with theoretical predictions, the effects of experimental resolution 
can be allowed for either by smearing the predictions, or by unfolding the data. The latter is a more difficult
procedure. I therefore recommend making available the experimental smearing matrix, rather than unfolding. 
There are few circumstances in which unfolding is really preferable.

\subsection{Statisticians}
For all of these discussions, it was extremely valuable having Statisticians at the meetings. At Kashiwa, we had
Sara Algeri, Michael Betancourt, David van Dyk. and Shiro Ikeda. 
Those at Fermilab were David van Dyk, Todd Kufner, Michael Kuusela, Richard Lockhart, Xiao Li Meng,
and Aixin Tan. As well as their presentations, it was most valuable having them available for informal discussions 
during breaks between the sessions.

\section{Conclusions} 
A post-Workshop survey showed that most participants felt that such meetings were worth-while, and would favour 
having more at a frequency of around 20 nanohertz. Smaller meetings devoted to statistical issues in specific analyses
(e.g. the neutrino mass hierarchy) could also be useful.

Another request was for more introductory material than there was time for at these Workshops. LL and Lorenzo Moneta subsequently gave a course of lectures plus computing practicals at CERN\cite{Moneta_CERN} and at IPMU\cite{Moneta_IPMU}.

Some of the large Collaborations at colliders have their own Statistics Committees for dealing with statsitical
issues arising in their own experiments. Neutrino experiments are in general too small for this, but a 
possibility would be to have a single forum for discussing statistical problems for all neutrino analyses.

It was gratifying to see that the Neutrino2016 Conference programme contained an invited talk by a Statistician\cite{DvD}.

\end{document}